\documentclass[a4paper,11pt]{article}
\usepackage{pos}
\usepackage{amsmath,graphicx,multirow,tabularx,physics}
\usepackage{davitex,slashed,listings}

\newcommand{\hp}{\hat{p}}

\newcommand{\he}{\hat{\epsilon}}

\newcommand{\hmu}{\muh}

\title{Isothermal and isentropic speed of sound in (2+1)-flavor QCD at non-zero baryon chemical potential}
\ShortTitle{Speed of Sound at non-zero chemical potential}

\author*[a,b,1]{D. A. Clarke}

\affiliation[a]{Department of Physics and Astronomy, University of Utah, \\
Salt Lake City, Utah, United States}
\affiliation[b]{Fakult\"at f\"ur Physik, Universit\"at Bielefeld,\\
Bielefeld, Germany}
\note{For the HotQCD collaboration.}
\emailAdd{clarke.davida@gmail.com}

\abstract{
Recently interest in calculations of the speed of sound in QCD under conditions like constant 
temperature $c^2_T$ or constant 
entropy per net baryon number $c^2_s$ arose in the discussion of experimental results coming from 
heavy ion experiments. 
It has been stressed that the former in particular is closely related to higher order cumulants of 
conserved charge fluctuations 
that are calculated in lattice QCD.
We present here results on $c^2_T$ and $c^2_s$ and compare results at vanishing strangeness chemical
potential and vanishing net 
strangeness number with hadron resonance gas model calculations. We stress the difference of both 
observables at low temperature
arising from the light meson sector, which does not contribute to $c^2_T$.
}

\FullConference{%
 The 39th International Symposium on Lattice Field Theory, LATTICE2022
  26th-30th July, 2022
  Bonn, Germany
}

\begin{document}
\maketitle

\section{Introduction}\label{sec:intro}

The isentropic speed of sound and isothermal speeds of sound are given by, respectively,
\begin{equation}
  c_s^2=\left(\pdv{p}{\epsilon}\right)_{s/n_B}~~~~~\text{and}~~~~~
  c_T^2=\left(\pdv{p}{\epsilon}\right)_T,
\end{equation}
where $p$ is the pressure, $\epsilon$ is the energy density, $s$ is the entropy density,
$n_B$ is the net baryon-number density, and $T$ is the temperature.
It is one of many bulk thermodynamic observables useful for characterizing
strongly interacting matter. For instance in the simple
Bjorken flow model, assuming a constant $c_s^2$, one can show~\cite{Bjorken:1982qr}
that the energy density will decrease with proper time $\tau$ as
$\tau^{-(1+c_s^2)}$.
In the context of heavy ion collisions (HIC), the system cools with longitudinal expansion
of the fireball according to $c_s^2$ in this picture.
Also in the context of HIC, it can be used to look out
for a long-lived fireball, which may coincide with a softest point
where the pressure-to-energy-density ratio,
and hence $c_s^2$, attains a minimum~\cite{Hung:1994eq}.
The isothermal speed of sound may also be of interest in the context of HIC,
as a new method to estimate $c_T^2$ in HIC has been recently suggested
in Ref.~\cite{Sorensen:2021zme}.
In the context of neutron stars, $c_s^2$ is interesting since the relationship
between the star masses and radii is influenced by how $c_s^2$ changes with
$n_B$~\cite{Ozel:2016oaf}.
This context is particularly interesting, since some situations
may suggest or require $c_s^2$ exceed its conformal limit
$1/3$~\cite{McLerran:2018hbz,Drischler:2020fvz,Fujimoto:2022ohj}.

With these applications in mind, it is worthwhile to revisit lattice investigations
of the speed of sound.
The speed of sound has been extensively studied at $\mu_B=0$ on the
lattice~\cite{Borsanyi:2010cj,HotQCD:2014kol,Borsanyi:2013bia}.
Here we extend these results to obtain a first calculation of $c_s^2$ at finite
baryon, electric charge, and strangeness chemical potentials
$\mu_B$, $\mu_Q$, and $\mu_S$ on the lattice. In order to obtain observables that are
functions of $\mu_B$ and $T$ only, and in order to target physics of
interest to HIC, we introduce two constraints
\begin{equation}\label{eq:constraints}
    n_S =0 ~~~~~~~\text{and}~~~~~~~n_Q/n_B=r,
\end{equation}
where $n_S$ and $n_Q$ are the net strangeness and electric charge densities, and $r=0.4$ or $0.5$
corresponding respectively to collisions at the Relativistic Heavy Ion Collider (RHIC)
and the isospin-symmetric case.

Thermodynamic observables including $c_s^2$ calculated at $r=0.5$ have been studied 
extensively by us in a recent publication~\cite{EoS}. This extends previous $6\nth$-order
results~\cite{Bazavov:2017dus} up to $8\nth$-order in the pressure series. In these proceedings, 
we supplement our most recent results with a calculation
of $c_T^2$ at $\mu_Q=\mu_Q=0$ and extend speed of sound results on lines of constant $s/n_B$
to include $r=0.4$, which is similar to the RHIC scenario. We confirm that differences in
$c_s^2$ and lines of constant $s/n_B$ arising from this change in $r$ are negligible.
For $c_T^2$, we will introduce instead the constraint $\mu_Q=\mu_S=0$. While less directly
relevant to HIC, this situation has $\mu_Q=0$ in common with $r=0.5$ and has the advantage of
especially simple expressions for $c_T^2$.

\section{Strategy of calculations}

The general strategy starts with finding $p$.
Once we have $p$, we can derive all other quantities from basic thermodynamic relations.
For temperatures near and above $\Tpc$ we use lattice QCD; near and below $\Tpc$ we use
the hadron resonance gas (HRG) model.

\subsection{Lattice QCD}
For convenience, we introduce dimensionless variables
$\hat{X}\equiv XT^{-k}$ with $k\in\Z$ chosen such that $\hat{X}$
is dimensionless. Thermodynamic observables are determined using the
Taylor expansion approach, i.e. we expand
\begin{equation}
\hp =\frac{1}{VT^3}\log\ZQCD(T,V,\hmu_B,\hmu_Q,\hmu_S) 
=\sum_{i,j,k=0}^\infty
\frac{\chi_{ijk}^{BQS}}{i!j!k!} \hmu_B^i \hmu_Q^j \hmu_S^k,
\label{Pdefinition}
\end{equation}
with expansion coefficients
\begin{equation}
\chi_{ijk}^{BQS}\equiv \chi_{ijk}^{BQS}(T) = 
\frac{\partial \hp}{\partial\hmu_B^i \partial\hmu_Q^j \partial\hmu_S^k} \;\Bigg|_{\hmu=0}.
\label{suscept}
\end{equation}
Imposing our constraints~\eqref{eq:constraints}
renders $\muh_Q$ and $\muh_S$ functions of $\muh_B$ and $T$, and hence we can 
reorganize~\footnote{The convergence of this series in $\muh_B$ was analyzed in
Ref.~\cite{Bollweg:2022rps}. There, it was argued that 
for $T\ge 130$~MeV, the $\hp$ series is reliable for $\muh_B\le 2.5$.
A similar analysis for $r=0.4$ delivers the same range of applicability~\cite{jishnu}.}
$\hp$ as
\begin{equation}
    \hp=P_0+\sum_{k=1}^{\infty} P_{2k}(T) \hmu_B^{2k}
\end{equation}
For more details on our implementation of constraints, see e.g. Ref.~\cite{Bazavov:2020bjn,jishnu,EoS}.

Perhaps the most straightforward strategy\footnote{Another strategy is given in
Appendix C of Ref.~\cite{EoS}.} to obtain $c_s^2$ on the lattice, and the one that
we employ here, is to use
\begin{equation}\label{eq:cs2numerical}
     c_s^2 \equiv
      c^2_{\vec{X}}
      =\left(
    \frac{\partial p}{\partial \epsilon}\right)_{\vec{X}} 
    =\frac{\left(
    \partial p/\partial T\right)_{\vec{X}}}{\left(\partial \epsilon/\partial T\right)_{\vec{X}}},
\end{equation}
where $\vec{X}\equiv(s/n_B,r,n_S)$. In this strategy, one takes numerical $T$-derivatives
of $p(T)$ and $\epsilon(T)$ that were determined along the line of constant physics $\vec{X}$.

When $T$ is held fixed, one can proceed analytically a bit further in a relatively straightforward
manner through Taylor expansion. In particular one has in this case
\begin{equation}
    c_T^2 = \left(\frac{\partial p}{\partial \epsilon}\right)_T =
    \left(\frac{\partial p}{\partial \hmu_B}\right)  \left(\frac{\partial \epsilon}{\partial \hmu_B}\right)^{-1}.
\end{equation}
When $\hmu_Q=\hmu_S=0$, the relationship between Taylor coefficients of $\hp$ and $\he$ become
especially simple, and one eventually finds
\begin{equation}\label{eq:invcT2}
    c_T^{-2} -3 =\frac{2 P'_2 \hmu_B + 4 P'_4 \hmu_B^3  + \order{\hmu_B^5}}{2P_2 \hmu_B + 4 P_4 \hmu_B^3  + \order{\hmu_B^5}}
    ~~~~~~\text{or}~~~~~~
    c_T^{-2} = 3 + \frac{P'_2}{P_2} + \sum_{k} c_{T,2k} \hmu_B^{2k}.
\end{equation}
Using the notation $X' =T {\rm d X}/{\rm d} T $, we
get for the expansion coefficients
\begin{equation}\label{eq:cT2coeff}
c_{T,2} = \left(\frac{P_4}{P_2}\right)',~~~~
c_{T,4} = 4 \left(\frac{P_4}{P_2}\right) \left(\frac{P_4}{P_2}\right)' + 3 \left(\frac{P_6}{P_2}\right)',~~~~... 
\end{equation}

\subsection{Hadron resonance gas}
In the HRG model, we work in a phase where quarks are confined so that the only degrees of
freedom are hadronic bound states. Hence this model is expected to be valid
up to roughly $\Tpc$. A non-interacting, quantum, relativistic gas eventually
delivers for particle species $i$
\begin{equation}\label{eq:HRG}
  \frac{p_i}{T}=\frac{m_i^2g_iT}{2\pi^2}\sum_{k=1}^\infty\frac{\eta_i^{k+1}z_i^k}{k^2}
                         K_2\left(\frac{m_ik}{T}\right),~~~~ ~~~~
  z_i\equiv e^{\,\muh_BB_i+\muh_QQ_i+\muh_SS_i},
\end{equation}
where $m_i$ is the species' mass, $g_i$ is its degeneracy factor, $\eta_i=\pm1$ for boson/fermion
statistics, and $K_2$ is the modified Bessel function\footnote{$K_2$ is exponentially suppressed,
so in practice we calculate eq.~\eqref{eq:HRG} numerically by dropping all terms with $k>20$. For the same
reason, we neglect states with masses larger than the kaon.} of the $2\nd$ kind.
The total $p$ is then found by summing over all 
known\footnote{We use the QMHRG2020 list of hadron resonances \cite{Bollweg:2021vqf}.} states.

In the special case $\hmu_Q=\hmu_S=0$, one can derive a relatively simple form for the
isothermal speed of sound. This case is instructive to get some intuition about how the speed of
sound behaves, especially at low temperatures, and it moreover shares $\hmu_Q=0$ in common
with the $r=0.5$ case. One schematically has in this situation
\begin{equation}\begin{aligned}\label{eq:peHRG}
\hp   &= f_M(T) + f_B(T) \cosh(\hmu_B), \\
\he   &=  3f_M(T) + f'_M(T) + \left(3f_B(T)+ f'_B(T)\right) \cosh(\hmu_B),\\
\end{aligned}\end{equation}
where $f_M(T)$ and $f_B(T)$ are the mesonic and baryonic contributions, respectively.
Hence when taking a $\hmu_B$-derivative, $f_M$ drops out. This makes computing the isothermal
speed of sound especially\footnote{This works nicely since $\mu_B$ and $T$ are
independent control parameters, so one can straightforwardly take a partial derivative
of one while holding the other fixed. By contrast, derivatives on a line of fixed
$s/n_B$ are much more delicate.} straightforward:
\begin{equation}\label{eq:cT2HRG}
    c_T^2 =
    \left(\frac{\partial p}{\partial \hmu_B}\right)
    \left(\frac{\partial \epsilon}{\partial \hmu_B}\right)^{-1}
    = \frac{1}{3+f'_B(T)/f_B(T)},
\end{equation}
i.e. in an HRG, $c_T^2$ will be $\hmu_B$-independent.
This is in agreement with the $r=0.5$
expansion coefficients of $c_T^2$ given in eq.~\eqref{eq:cT2coeff}. To see this,
note that for an HRG in the Boltzmann
approximation, the expansion coefficients $P_{2k}$
are given by 
\begin{equation}
    P_{2k}=\frac{f_B(T)}{2k!}.
\end{equation} The ratios 
$P_{2k}/P_2$ are thus $T$-independent, which means the coefficients in
eq.~\eqref{eq:cT2coeff} vanish when applied to a $\muh_Q=\hmu_S=0$ HRG.

To determine $c_s^2$ in HRG, one could use eq.~\eqref{eq:cs2numerical}.
While this is quite successful for large $s/n_B$, which corresponds\footnote{The
$s(\hmu_B)$ expansion has a nonzero leading term $s_0$, while $n_B(\hmu_B)$ leads at
$\order{\hmu_B}$. Thus the limit $\hmu\to0$ corresponds to $s/n_B\to\infty$.}
to small $\hmu_B$, we found it had numerical difficulties for $s/n_B\ltsim10$.
Instead, we use here Appendix C of Ref.~\cite{EoS}, which while more elaborate to implement,
increases numerical stability by circumventing the numerical $T$-derivatives.
We find exact agreement between both approaches for $s/n_B\gtsim400$, while the
second approach allows us to compute $c_s^2$ for $s/n_B\ltsim10$ more reliably.

\section{Computational setup}\label{sec:setup}

We use high-statistics data sets for
$(2+1)$-flavor QCD with degenerate light quark masses
$m_u=m_d\equiv m_l$ and a heavier strange quark mass $m_s$.
These data sets were generated with the
HISQ action using $\simulat$~\cite{Bollweg:2021cvl}
and have been presented in previous HotQCD studies~\cite{Bazavov:2017dus,Bollweg:2022rps}. 

For $T<180$~MeV, the speed of sound is extracted from continuum-extrapolated 
data\footnote{For details on our continuum extrapolation, see Ref.~\cite{EoS}.} from
$N_\tau=8$, $12$, and $16$ lattices with $m_s/m_l=27$, which is the physical value.
For $T> 180$~MeV, we use data~\cite{Bazavov:2017dus}
with slightly heavier\footnote{This is known to have a negligible effect on the results~\cite{Bazavov:2011nk}.} 
light quarks, $m_s/m_l=20$.
In all cases results have been obtained on
lattices with aspect ratio $N_\sigma/N_\tau=4$. 

We are often interested in the behavior of observables near the pseudocritical temperature
$\Tpc$. When indicated on figures, we take $\Tpc =156.5(1.5)$~MeV from Ref.~\cite{HotQCD:2018pds}.
$\Tpc(\muh_B)$ curves use the $\order{\hmu_B^2}$ expansion
\begin{equation}
    \Tpc(\hmu_B)=\Tpc(0) \left( 1-\kappa_2^{B}\hmu_B^2 +\order{\hmu_B^4}\right)
\end{equation}
using curvature coefficient $\kappa_2^B=0.016$ for $r=0.5$ and $\kappa_2^B=0.012$ for $r=0.4$.

The \texttt{AnalysisToolbox}~\cite{toolbox} is used to facilitate HRG calculations and bootstrapping. Statistical
uncertainty in all figures is represented by bands and is calculated through bootstrap resampling, unless otherwise
stated. Central values are returned as the median, with the lower and upper error bounds given by the 32\% and 68\%
quantiles, respectively. If needed, spline interpolations are cubic with evenly spaced knots, and temperature
derivatives of lattice QCD data are calculated by fitting the temperature dependence with a spline, then calculating
the derivative of the spline numerically.

\section{Results}\label{sec:results}

\begin{figure}
\centering
\includegraphics[width=0.495\linewidth]{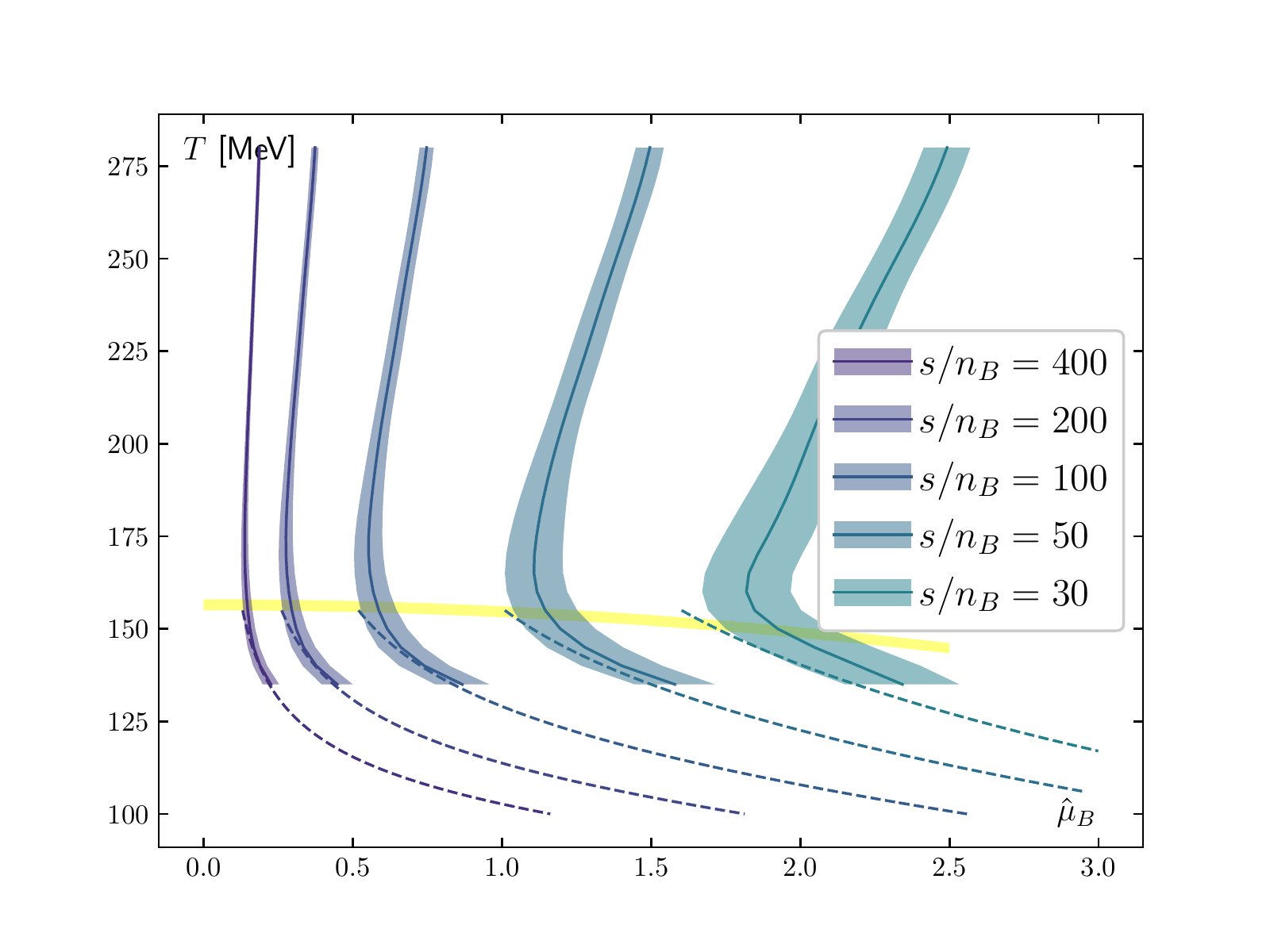}
\includegraphics[width=0.495\linewidth]{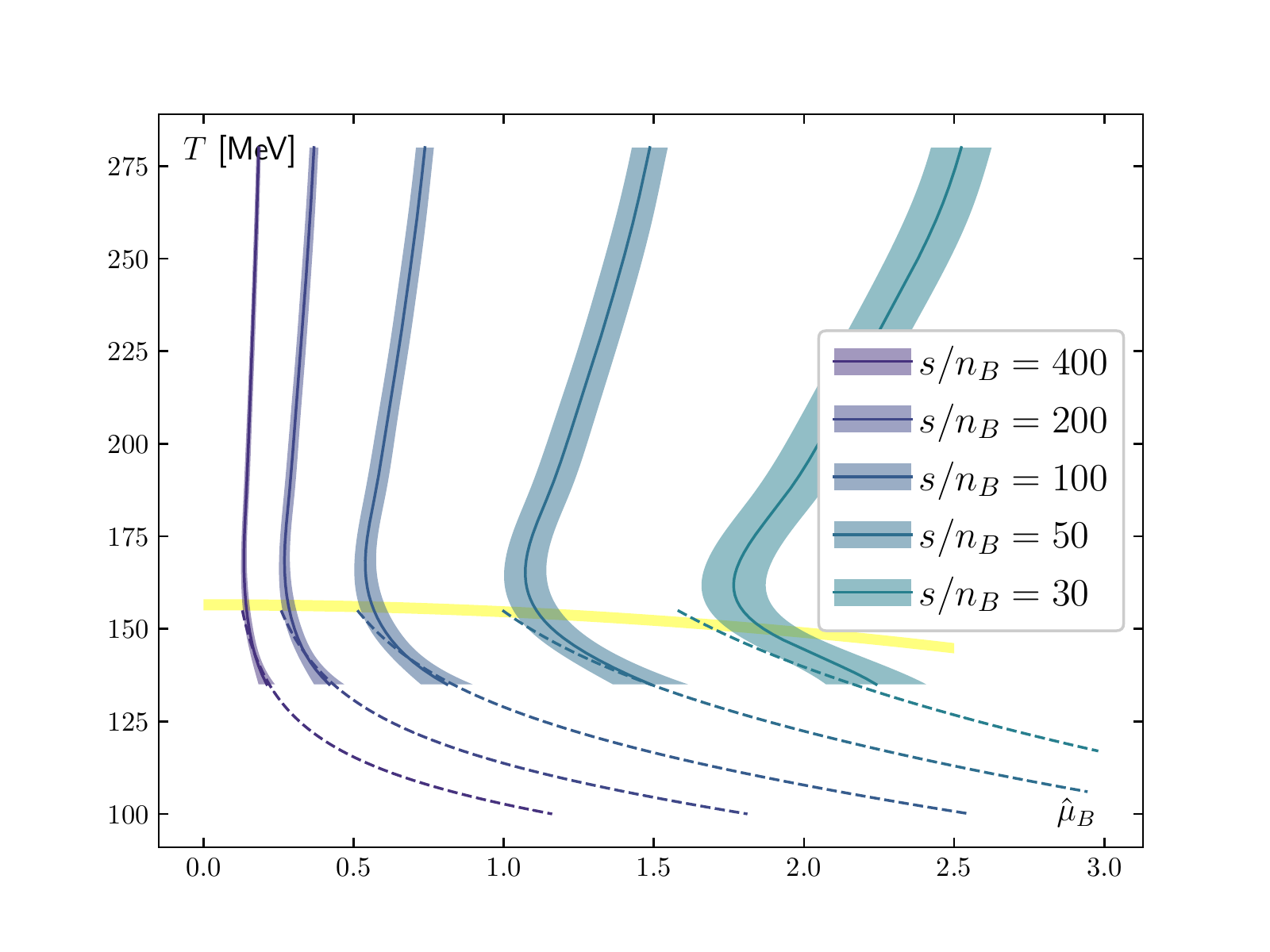}
\caption{Lines of constant entropy per baryon number in the $T$-$\hmu_B$
plane for $r=0.4$ (left) and $r=0.5$ (right). 
Solid bands indicate results obtained by numerically solving
$s/n_B$ derived from the $\order{\hmu_B^6}$ pressure series for $\hmu_B$. 
Dashed lines indicate QMHRG2020 model calculations.
The yellow band indicates $\Tpc(\hmu_B)$.}
\label{fig:snBLCP}
\end{figure}

\begin{figure}
\centering
\includegraphics[width=0.495\linewidth]{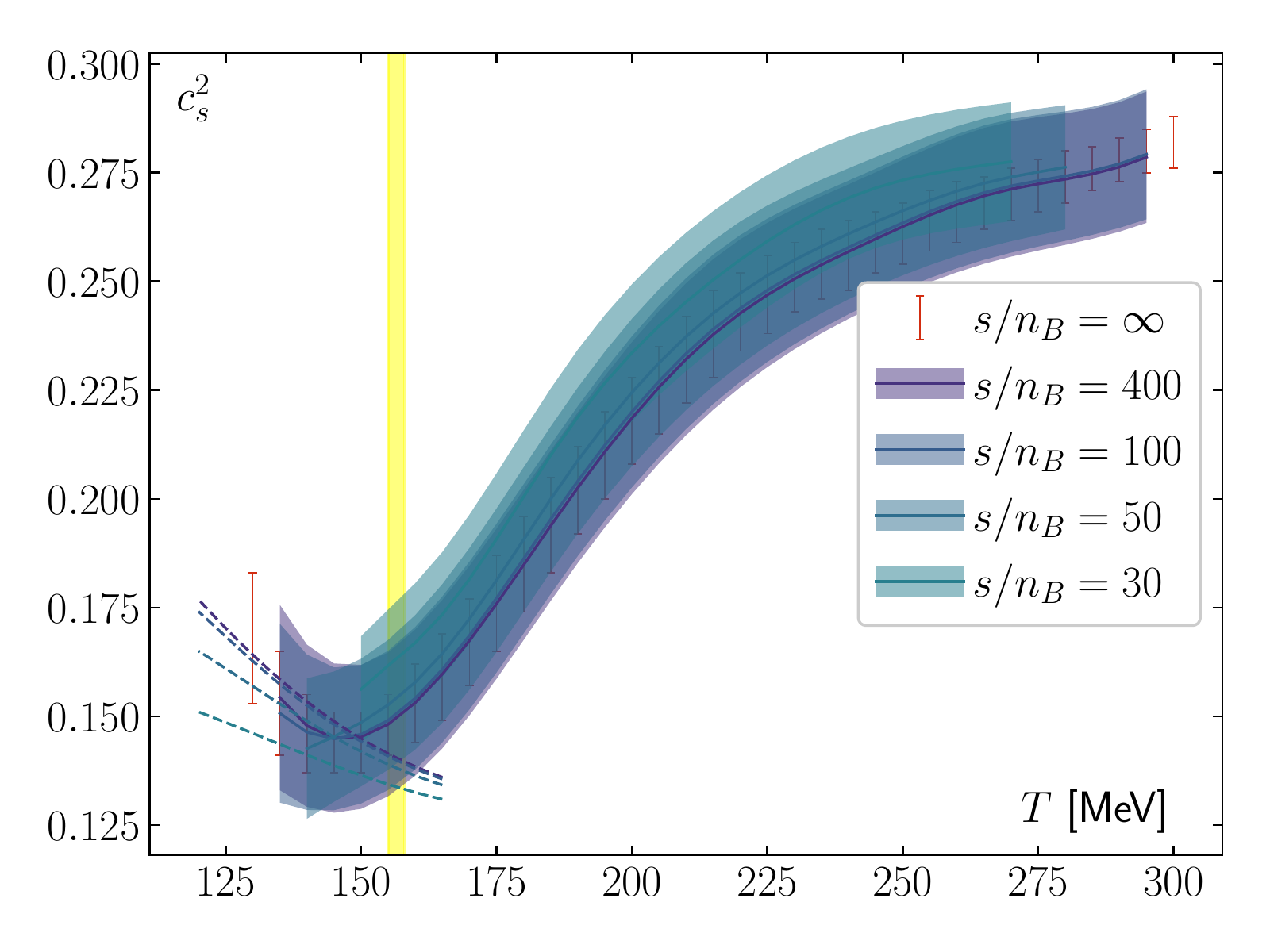}
\includegraphics[width=0.495\linewidth]{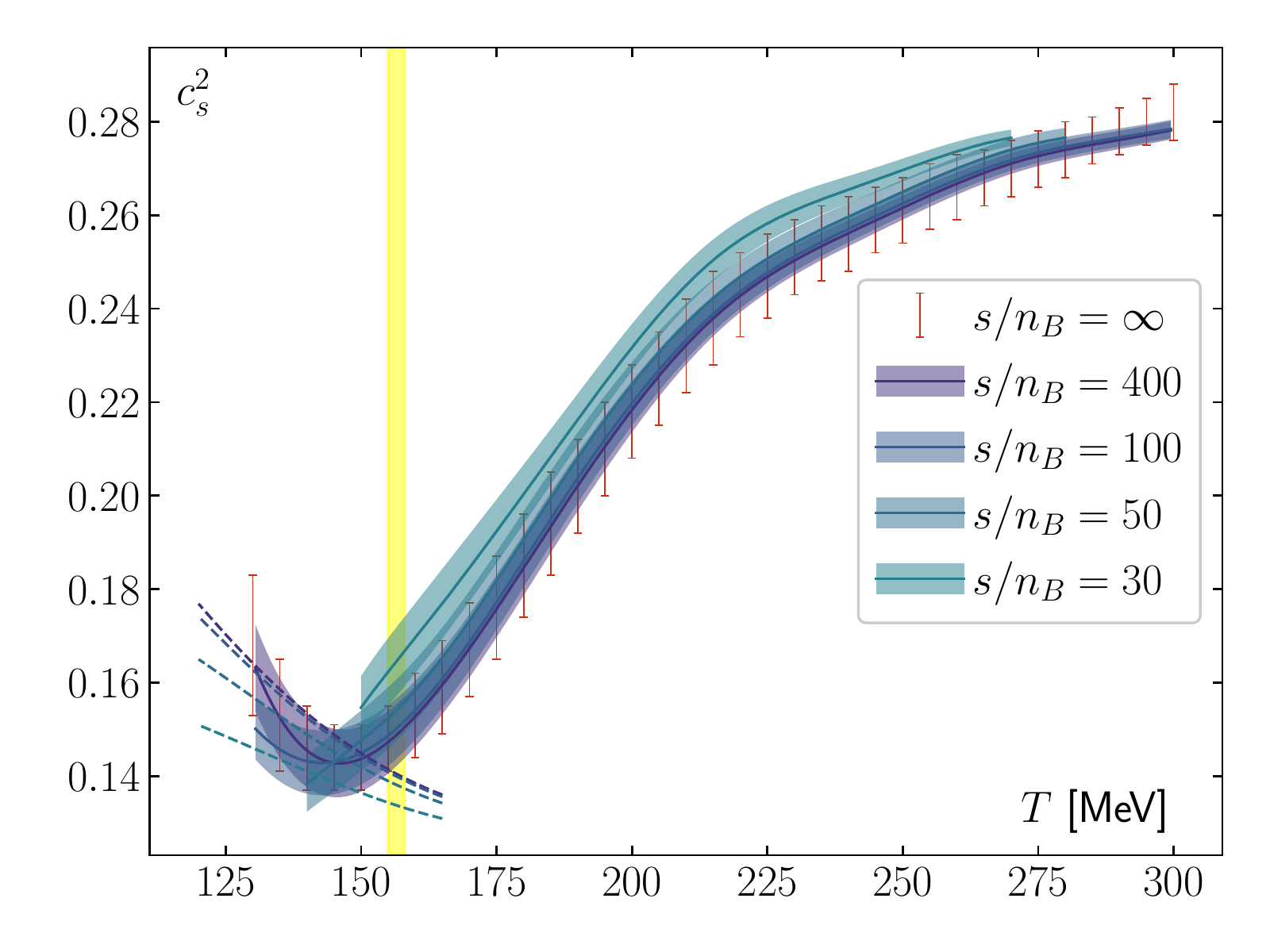}
\caption{Isentropic speed of sound versus temperature for strangeness-neutral 
matter with $r=0.4$ (left) 
and $r=0.5$ (right). Dashed lines at low temperatures indicate QMHRG2020 model calculations,
while the yellow band indicates $\Tpc$.
$s/n_B=\infty$ data taken from Ref.~\cite{HotQCD:2014kol}.}
\label{fig:cs2}
\end{figure}

\begin{figure}
\centering
\includegraphics[width=0.49\linewidth]{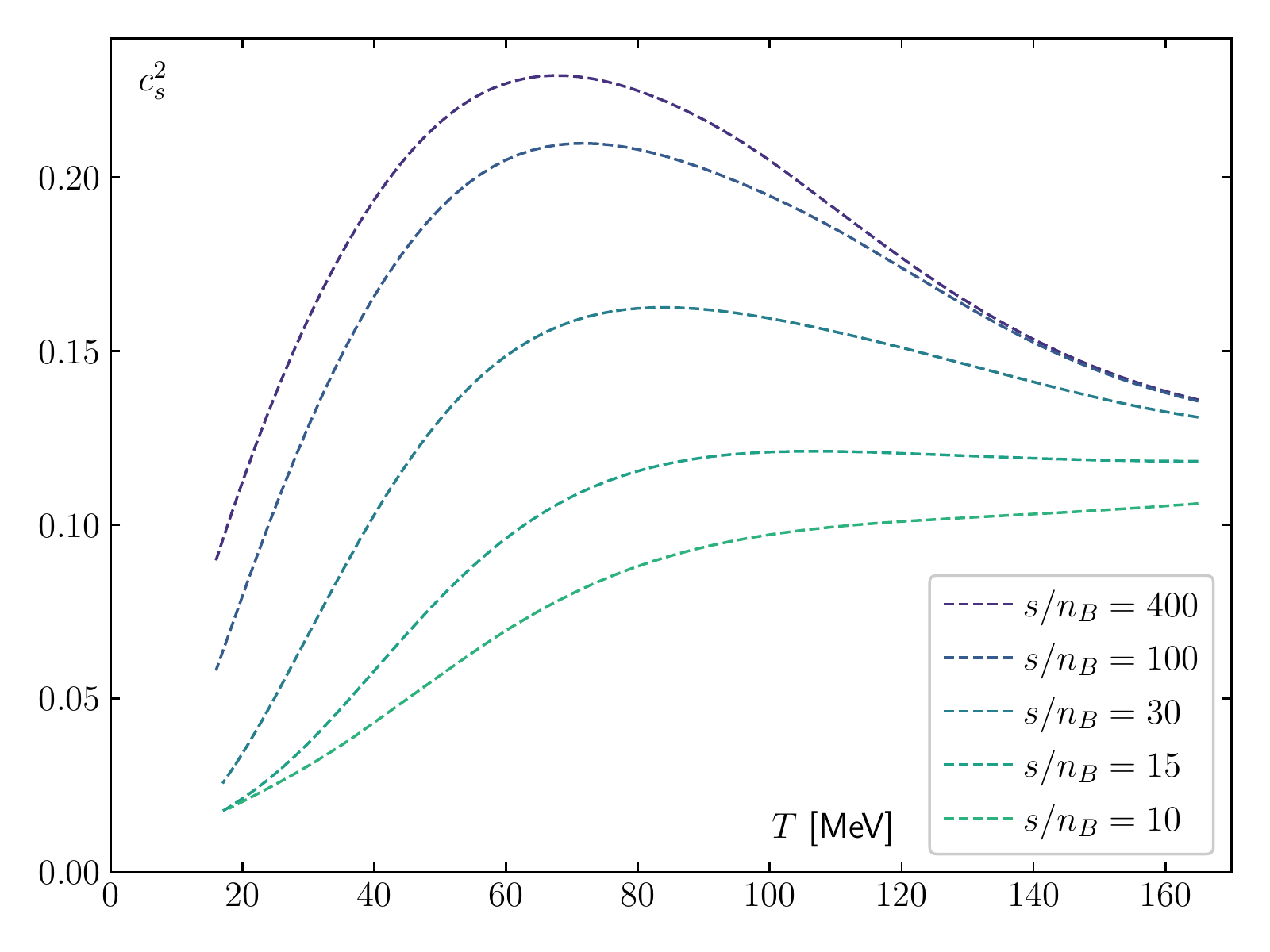}
\includegraphics[width=0.49\linewidth]{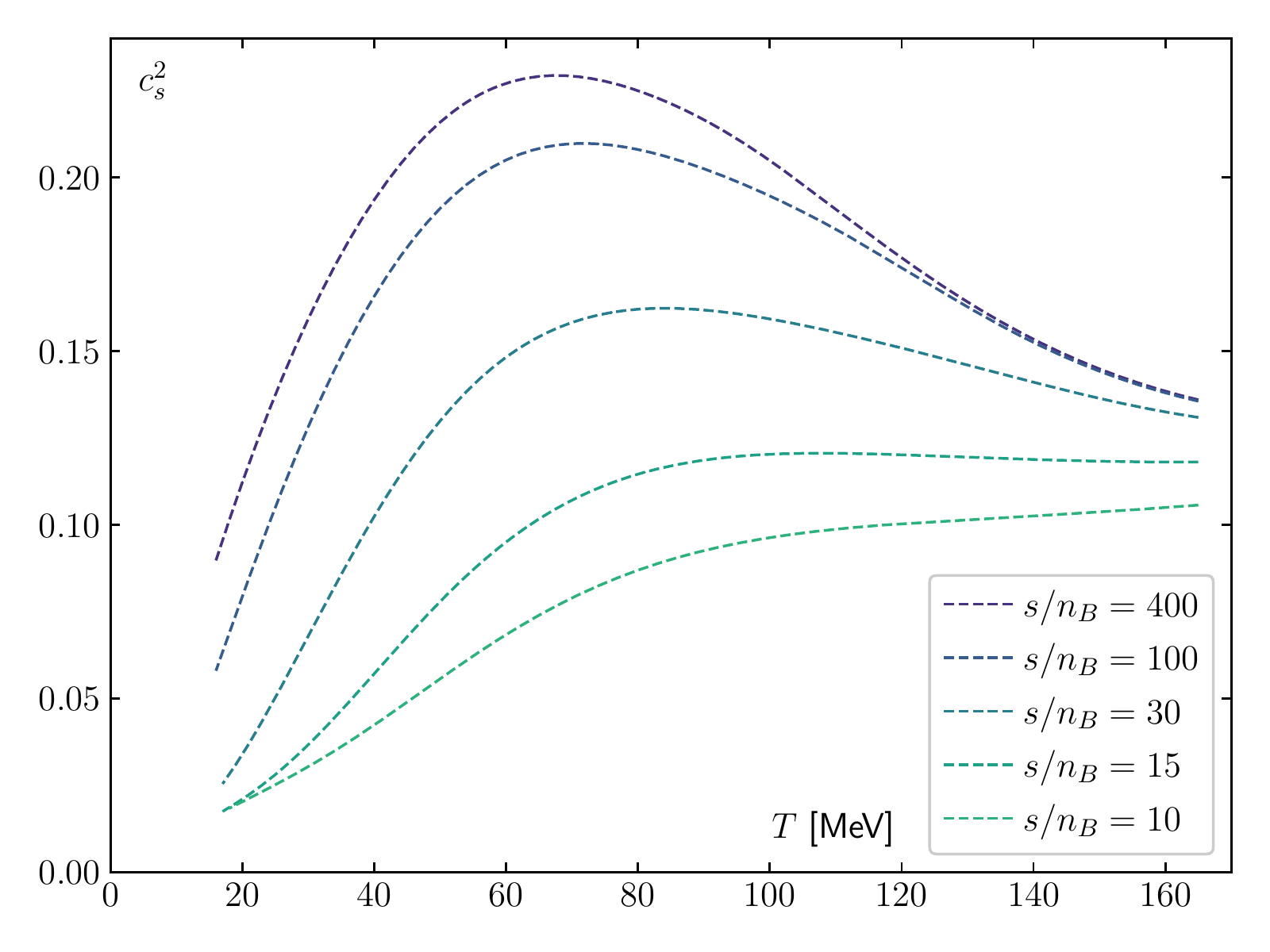}
\caption{Isentropic speed of sound versus temperature for strangeness-neutral matter
with $r=0.4$ (left) and $r=0.5$ (right) from QMHRG2020 model calculations.}
\label{fig:cs2HRG}
\end{figure}

Results for $c_s^2$ are computed along lines of constant $s/n_B$, which are
depicted for both the $r=0.4$ and $r=0.5$ cases in Fig.~\ref{fig:snBLCP}. 
We examine $400\leq s/n_B\leq30$, which very roughly
corresponds to the $s/n_B$ range covered by
BES-II at RHIC for beam energies $7.7~{\rm GeV}\ltsim \sqrt{s_{_{NN}}} \ltsim 200~{\rm GeV}$.
We find good agreement with HRG below $\Tpc$.
For Figs.~\ref{fig:snBLCP}, \ref{fig:cs2}, and \ref{fig:cs2HRG}, the
behavior between the $r=0.4$ and $r=0.5$ cases is qualitatively the same
and quantitatively very close, i.e. we verify that differences in these
observables due to deviations from the isospin-symmetric case are quite small.
Error bars for the $r=0.4$ case may be larger, since one introduces an
error in $\hmu_Q(\hmu_B)$, which is otherwise exactly zero in the $r=0.5$ case.

In Fig.~\ref{fig:cs2} we show our results for $c_s^2$ against $T$ for both
$r=0.4$ and $r=0.5$. Fig.~\ref{fig:cs2HRG} shows the HRG results
down to about $T=20$ MeV.
In general one finds only mild quantitative differences with changing $s/n_B$
above $\Tpc$.
We find good agreement between lattice results and HRG below $\Tpc$.
Near $\Tpc$, one finds a dip in the lattice data for $s/n_B\geq 100$.
Using both lattice and HRG results, one expects a dip also down
to at least $s/n_B=30$. This dip location roughly corresponds to the location
of the $p/\epsilon$ minimum, i.e. the softest point mentioned in the
introduction, which one can also verify directly using
our $p$ and $\epsilon$ data~\cite{EoS}. This gives yet another indication of
the existence of a crossover at all chemical potentials examined in this study.

\begin{figure}
\centering
\includegraphics[width=0.49\linewidth]{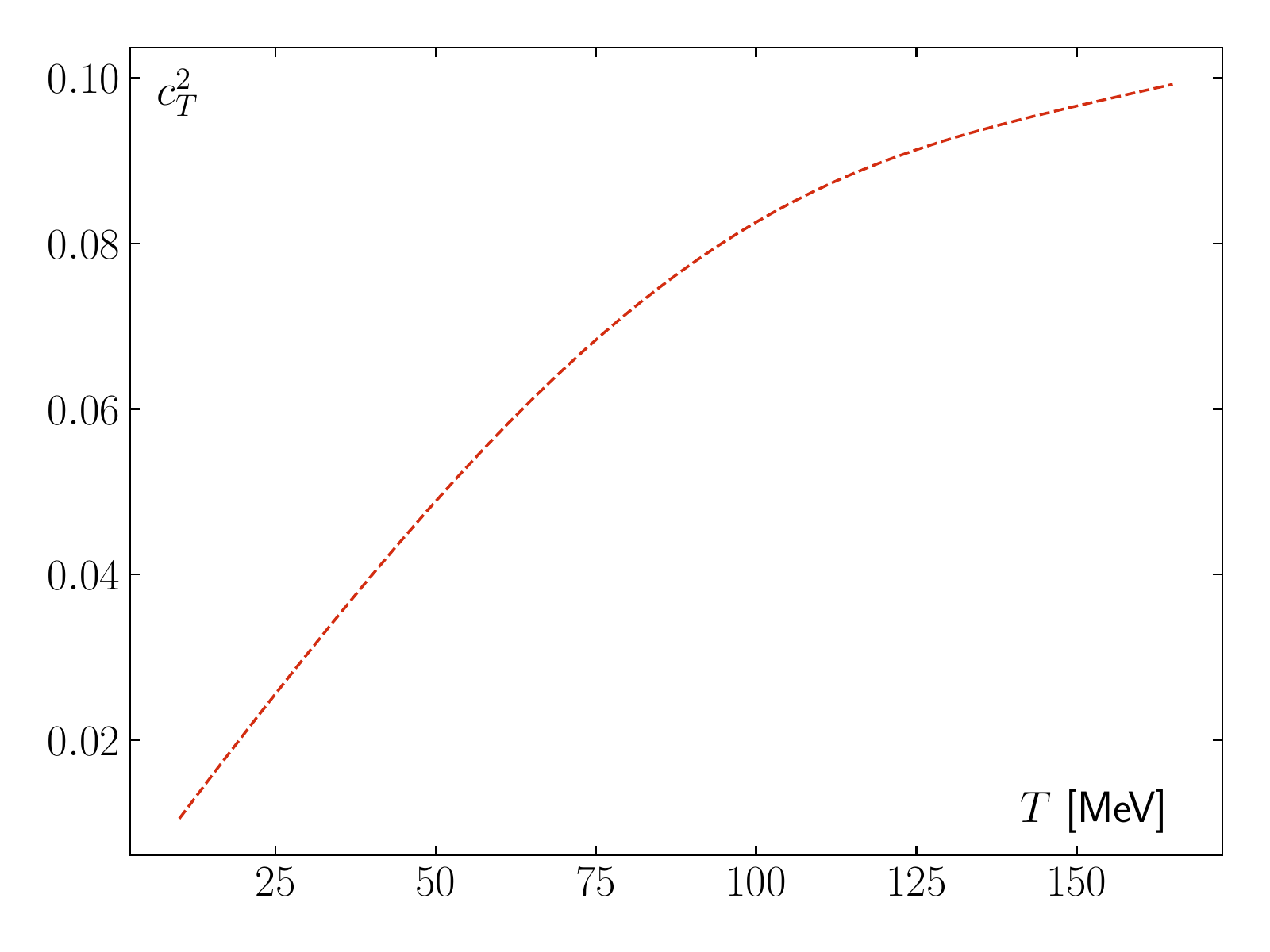}
\includegraphics[width=0.49\linewidth]{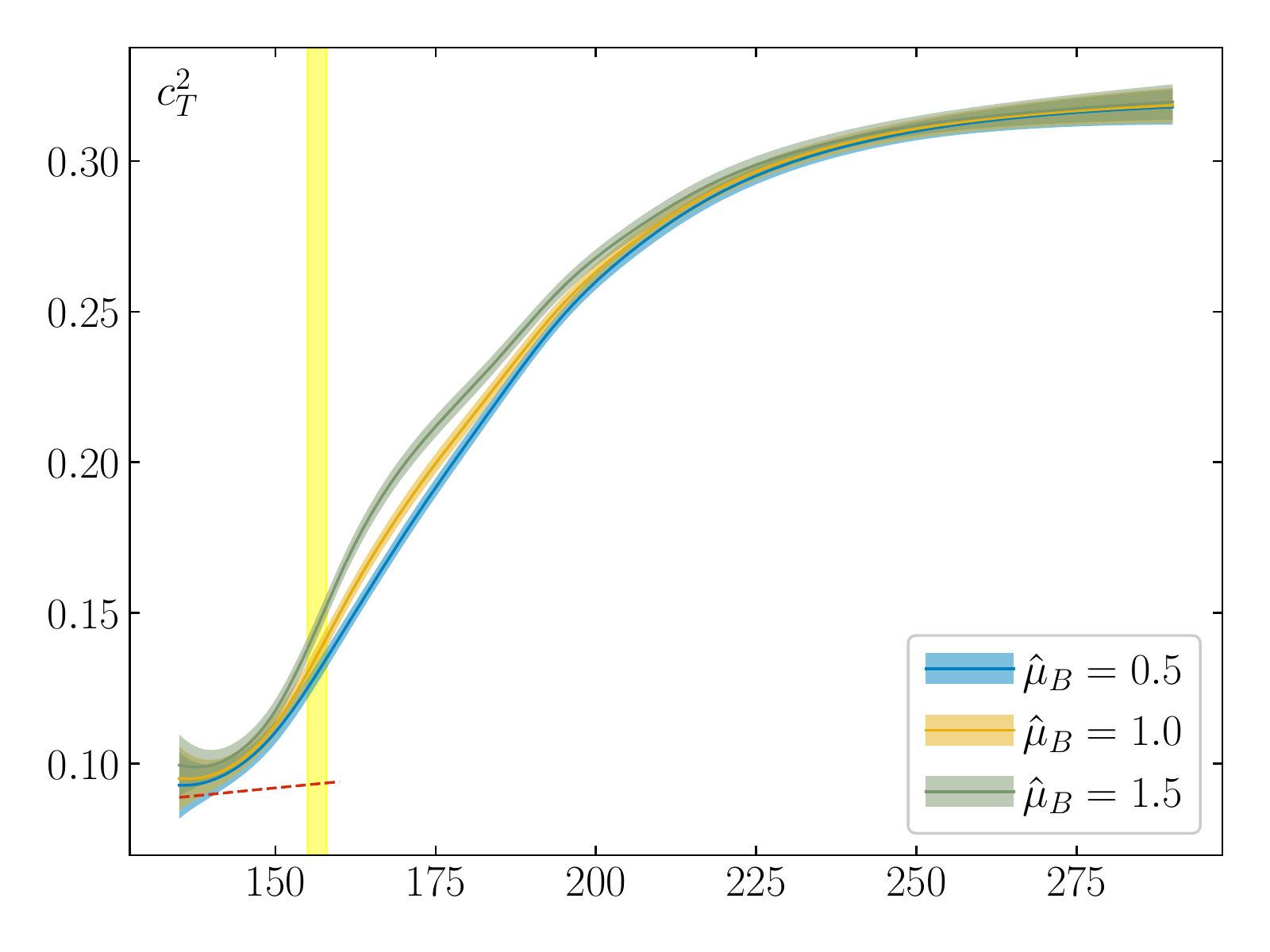}
\caption{Isothermal speed of sound for strangeness-neutral matter. {\it Left}:
         QMHRG2020 calculation at $\hmu_Q=\hmu_s=0$. {\it Right}: Lattice data
         at $r=0.5$. Error bands come from error propagation. The yellow band indicates $\Tpc$,
         and the red, dashed line indicates the HRG curve from the left figure.}
\label{fig:cT2}
\end{figure}

Turning to the QMHRG2020 results shown in Fig.~\ref{fig:cs2HRG}, we see a peak
in $c_s^2$ that decreases with decreasing $s/n_B$. Somewhere in the vicinity $s/n_B\in[10,15]$, 
the peak has vanished, and $c_s^2$ increases monotonically with $T$ up to 165~MeV.
The $c_s^2$ curves at $r=0.4$ and $r=0.5$ seem to approach $c_T^2$ as $s/n_B\to0$
with particularly close agreement at the lowest calculated $T$. We reiterate that
we only have $c_T^2$ data at $\hmu_Q=\hmu_S=0$, which is a somewhat different
situation than both $r=0.4$ and $r=0.5$. This precludes an unambiguous direct comparison.

From eq.~\eqref{eq:peHRG} and \eqref{eq:cT2HRG} we see that $c_T^2$ is insensitive to mesons.
We will use this as a starting point to understand the weakening of the peak in $c_s^2$.
In the massless limit, one expects from eq.~\eqref{eq:HRG} that $c_s^2$ and $c_T^2$ will
be 1/3 at all $T$. Continuing this behavior to small $m$, one expects that small masses
have the tendency to pull speed of sound curves up toward 1/3. The isentropic speed of sound,
which feels the mesonic sector, but should approach 0 at low $T$, therefore develops a peak.
By contrast $c_T^2$ at $\hmu_Q=\hmu_S=0$ is insensitive to mesons, so it has no tendency to be 
pulled to 1/3. In Fig.~\ref{fig:cT2} (right), we show a lattice determination of $c_T^2$
at $r=0.5$ using eq.~\eqref{eq:invcT2}. Despite the slight difference in external conditions, it
agrees well with HRG at low $T$, and it rapidly approaches the ideal gas limit 1/3 at high $T$.

As a closing remark, we mention that our results for the speed of sound are in rough 
qualitative agreement
with various model calculations, for instance PNJL and NJL
models~\cite{Ghosh:2006qh,Marty:2013ita,Deb:2016myz,Motta:2020cbr,Zhao:2020xob};
the quark-meson coupling
model~\cite{Schaefer:2009ui,Abhishek:2017pkp};
the field correlator
method~\cite{Khaidukov:2018lor,Khaidukov:2019icg};
and the quasiparticle method~\cite{Mykhaylova:2020pfk}.

\section{Conclusion and outlook}\label{sec:outlook}

We presented a first lattice calculation of $c_s^2$ and $c_T^2$ at finite chemical potential.
The dip in $c_s^2$ near $\Tpc$, or equivalently its peak at lower $T$,
can be understood through its sensitivity to light meson states. For all results we find a 
negligible difference between $r=0.4$ and $r=0.5$.
Our results for $c_s^2$ are qualitatively in agreement with model calculations. 
Finally we note that the strategy of Appendix C in Ref.~\cite{EoS} works
quite successfully for $c_s^2$, and hope to extend it to 
other thermodynamic observables. 

\section*{Acknowledgements}

D. C. was funded by the Deutsche Forschungsgemeinschaft (DFG, German Research 
Foundation) - Project numbers 315477589-TRR 211 and 
the ``NFDI 39/1" for the PUNCH4NFDI consortium.
This research used awards of computer time provided by:
(i) The INCITE program at Oak Ridge Leadership Computing Facility, a DOE
Office of Science User Facility operated under Contract No. DE-AC05-00OR22725;
(ii) The ALCC program at National Energy Research Scientific Computing Center,
a U.S. Department of Energy Office of Science User Facility operated under
Contract No. DE-AC02-05CH11231;
(iii) The INCITE program at Argonne Leadership Computing Facility, a U.S.
Department of Energy Office of Science User Facility operated under Contract
No. DE-AC02-06CH11357;
(iv) The USQCD resources at the Thomas Jefferson National Accelerator Facility.
This research also used computing resources made available through:
(i) a  PRACE grant at CINECA, Italy;
(ii) the Gauss Center at NIC-J\"ulich, Germany;
(iii) the GPU-cluster at Bielefeld University, Germany.

\bibliographystyle{JHEP}
\bibliography{bibliography}

\providecommand{\href}[2]{#2}\begingroup\raggedright\begin{thebibliography}{10}

\bibitem{Bjorken:1982qr}
J.~D. Bjorken, \emph{{Highly Relativistic Nucleus-Nucleus Collisions: The
  Central Rapidity Region}},
  \href{https://doi.org/10.1103/PhysRevD.27.140}{\emph{Phys. Rev. D} {\bfseries
  27} (1983) 140}.

\bibitem{Hung:1994eq}
C.~M. Hung and E.~V. Shuryak, \emph{{Hydrodynamics near the QCD phase
  transition: Looking for the longest lived fireball}},
  \href{https://doi.org/10.1103/PhysRevLett.75.4003}{\emph{Phys. Rev. Lett.}
  {\bfseries 75} (1995) 4003}
  [\href{https://arxiv.org/abs/hep-ph/9412360}{{\ttfamily hep-ph/9412360}}].

\bibitem{Sorensen:2021zme}
A.~Sorensen, D.~Oliinychenko, V.~Koch and L.~McLerran, \emph{{Speed of Sound
  and Baryon Cumulants in Heavy-Ion Collisions}},
  \href{https://doi.org/10.1103/PhysRevLett.127.042303}{\emph{Phys. Rev. Lett.}
  {\bfseries 127} (2021) 042303}
  [\href{https://arxiv.org/abs/2103.07365}{{\ttfamily 2103.07365}}].

\bibitem{Ozel:2016oaf}
F.~\"Ozel and P.~Freire, \emph{{Masses, Radii, and the Equation of State of
  Neutron Stars}},
  \href{https://doi.org/10.1146/annurev-astro-081915-023322}{\emph{Ann. Rev.
  Astron. Astrophys.} {\bfseries 54} (2016) 401}
  [\href{https://arxiv.org/abs/1603.02698}{{\ttfamily 1603.02698}}].

\bibitem{McLerran:2018hbz}
L.~McLerran and S.~Reddy, \emph{{Quarkyonic Matter and Neutron Stars}},
  \href{https://doi.org/10.1103/PhysRevLett.122.122701}{\emph{Phys. Rev. Lett.}
  {\bfseries 122} (2019) 122701}
  [\href{https://arxiv.org/abs/1811.12503}{{\ttfamily 1811.12503}}].

\bibitem{Drischler:2020fvz}
C.~Drischler, S.~Han, J.~M. Lattimer, M.~Prakash, S.~Reddy and T.~Zhao,
  \emph{{Limiting masses and radii of neutron stars and their implications}},
  \href{https://doi.org/10.1103/PhysRevC.103.045808}{\emph{Phys. Rev. C}
  {\bfseries 103} (2021) 045808}
  [\href{https://arxiv.org/abs/2009.06441}{{\ttfamily 2009.06441}}].

\bibitem{Fujimoto:2022ohj}
Y.~Fujimoto, K.~Fukushima, L.~D. McLerran and M.~Praszalowicz, \emph{{Trace
  anomaly as signature of conformality in neutron stars}},
  \href{https://arxiv.org/abs/2207.06753}{{\ttfamily 2207.06753}}.

\bibitem{Borsanyi:2010cj}
S.~Borsanyi, G.~Endrodi, Z.~Fodor, A.~Jakovac, S.~D. Katz, S.~Krieg et~al.,
  \emph{{The QCD equation of state with dynamical quarks}},
  \href{https://doi.org/10.1007/JHEP11(2010)077}{\emph{JHEP} {\bfseries 11}
  (2010) 077} [\href{https://arxiv.org/abs/1007.2580}{{\ttfamily 1007.2580}}].

\bibitem{HotQCD:2014kol}
{\scshape HotQCD} collaboration, \emph{{Equation of state in ( 2+1 )-flavor
  QCD}}, \href{https://doi.org/10.1103/PhysRevD.90.094503}{\emph{Phys. Rev. D}
  {\bfseries 90} (2014) 094503}
  [\href{https://arxiv.org/abs/1407.6387}{{\ttfamily 1407.6387}}].

\bibitem{Borsanyi:2013bia}
S.~Borsanyi, Z.~Fodor, C.~Hoelbling, S.~D. Katz, S.~Krieg and K.~K. Szabo,
  \emph{{Full result for the QCD equation of state with 2+1 flavors}},
  \href{https://doi.org/10.1016/j.physletb.2014.01.007}{\emph{Phys. Lett. B}
  {\bfseries 730} (2014) 99} [\href{https://arxiv.org/abs/1309.5258}{{\ttfamily
  1309.5258}}].

\bibitem{EoS}
D.~Bollweg, D.~A. Clarke, J.~Goswami, O.~Kaczmarek, F.~Karsch, S.~Mukherjee
  et~al., \emph{{Equation of state and speed of sound of (2+1)-flavor QCD in
  strangeness-neutral matter at non-vanishing net baryon-number density}},
  \href{https://arxiv.org/abs/2212.09043}{{\ttfamily 2212.09043}}.

\bibitem{Bazavov:2017dus}
A.~Bazavov et~al., \emph{{The QCD Equation of State to $\mathcal{O}(\mu_B^6)$
  from Lattice QCD}},
  \href{https://doi.org/10.1103/PhysRevD.95.054504}{\emph{Phys. Rev. D}
  {\bfseries 95} (2017) 054504}
  [\href{https://arxiv.org/abs/1701.04325}{{\ttfamily 1701.04325}}].

\bibitem{Bollweg:2022rps}
{\scshape HotQCD} collaboration, \emph{{Taylor expansions and Pad\'e
  approximants for cumulants of conserved charge fluctuations at nonvanishing
  chemical potentials}},
  \href{https://doi.org/10.1103/PhysRevD.105.074511}{\emph{Phys. Rev. D}
  {\bfseries 105} (2022) 074511}
  [\href{https://arxiv.org/abs/2202.09184}{{\ttfamily 2202.09184}}].

\bibitem{jishnu}
J.~Goswami, \emph{The isentropic equation of state of (2+1)-flavor qcd: An
  update based on high precision taylor expansion and padé-resummed expansion
  at finite chemical potentials}, {\emph{PoS} {\bfseries LATTICE2022} (2022)
  149} [\href{https://arxiv.org/abs/2212.10016}{{\ttfamily 2212.10016}}].

\bibitem{Bazavov:2020bjn}
{\scshape HotQCD} collaboration, \emph{{Skewness, kurtosis, and the fifth and
  sixth order cumulants of net baryon-number distributions from lattice QCD
  confront high-statistics STAR data}},
  \href{https://doi.org/10.1103/PhysRevD.101.074502}{\emph{Phys. Rev. D}
  {\bfseries 101} (2020) 074502}
  [\href{https://arxiv.org/abs/2001.08530}{{\ttfamily 2001.08530}}].

\bibitem{Bollweg:2021vqf}
{\scshape HotQCD} collaboration, \emph{{Second order cumulants of conserved
  charge fluctuations revisited: Vanishing chemical potentials}},
  \href{https://doi.org/10.1103/PhysRevD.104.074512}{\emph{Phys. Rev. D}
  {\bfseries 104} (2021) } [\href{https://arxiv.org/abs/2107.10011}{{\ttfamily
  2107.10011}}].

\bibitem{Bollweg:2021cvl}
D.~Bollweg, L.~Altenkort, D.~A. Clarke, O.~Kaczmarek, L.~Mazur, C.~Schmidt
  et~al., \emph{{HotQCD on multi-GPU Systems}},
  \href{https://doi.org/10.22323/1.396.0196}{\emph{PoS} {\bfseries LATTICE2021}
  (2022) 196} [\href{https://arxiv.org/abs/2111.10354}{{\ttfamily
  2111.10354}}].

\bibitem{Bazavov:2011nk}
{\scshape HotQCD} collaboration, \emph{{The chiral and deconfinement aspects of
  the QCD transition}},
  \href{https://doi.org/10.1103/PhysRevD.85.054503}{\emph{Phys. Rev. D}
  {\bfseries 85} (2012) 054503}
  [\href{https://arxiv.org/abs/1111.1710}{{\ttfamily 1111.1710}}].

\bibitem{HotQCD:2018pds}
{\scshape HotQCD} collaboration, \emph{{Chiral crossover in QCD at zero and
  non-zero chemical potentials}},
  \href{https://doi.org/10.1016/j.physletb.2019.05.013}{\emph{Phys. Lett. B}
  {\bfseries 795} (2019) 15}
  [\href{https://arxiv.org/abs/1812.08235}{{\ttfamily 1812.08235}}].

\bibitem{toolbox}
``{AnalysisToolbox: A set of Python tools for analyzing physics data, in
  particular targeting lattice QCD}.''
  \url{https://github.com/LatticeQCD/AnalysisToolbox}.

\bibitem{Ghosh:2006qh}
S.~K. Ghosh, T.~K. Mukherjee, M.~G. Mustafa and R.~Ray, \emph{{Susceptibilities
  and speed of sound from PNJL model}},
  \href{https://doi.org/10.1103/PhysRevD.73.114007}{\emph{Phys. Rev. D}
  {\bfseries 73} (2006) 114007}
  [\href{https://arxiv.org/abs/hep-ph/0603050}{{\ttfamily hep-ph/0603050}}].

\bibitem{Marty:2013ita}
R.~Marty, E.~Bratkovskaya, W.~Cassing, J.~Aichelin and H.~Berrehrah,
  \emph{{Transport coefficients from the Nambu-Jona-Lasinio model for
  $SU(3)_f$}}, \href{https://doi.org/10.1103/PhysRevC.88.045204}{\emph{Phys.
  Rev. C} {\bfseries 88} (2013) 045204}
  [\href{https://arxiv.org/abs/1305.7180}{{\ttfamily 1305.7180}}].

\bibitem{Deb:2016myz}
P.~Deb, G.~P. Kadam and H.~Mishra, \emph{{Estimating transport coefficients in
  hot and dense quark matter}},
  \href{https://doi.org/10.1103/PhysRevD.94.094002}{\emph{Phys. Rev. D}
  {\bfseries 94} (2016) 094002}
  [\href{https://arxiv.org/abs/1603.01952}{{\ttfamily 1603.01952}}].

\bibitem{Motta:2020cbr}
M.~Motta, R.~Stiele, W.~M. Alberico and A.~Beraudo, \emph{{Isentropic evolution
  of the matter in heavy-ion collisions and the search for the critical
  endpoint}}, \href{https://doi.org/10.1140/epjc/s10052-020-8218-x}{\emph{Eur.
  Phys. J. C} {\bfseries 80} (2020) 770}
  [\href{https://arxiv.org/abs/2003.04734}{{\ttfamily 2003.04734}}].

\bibitem{Zhao:2020xob}
Y.-P. Zhao, \emph{{Thermodynamic properties and transport coefficients of QCD
  matter within the nonextensive
  Polyakov\textendash{}Nambu\textendash{}Jona-Lasinio model}},
  \href{https://doi.org/10.1103/PhysRevD.101.096006}{\emph{Phys. Rev. D}
  {\bfseries 101} (2020) 096006}
  [\href{https://arxiv.org/abs/2004.14556}{{\ttfamily 2004.14556}}].

\bibitem{Schaefer:2009ui}
B.-J. Schaefer, M.~Wagner and J.~Wambach, \emph{{Thermodynamics of (2+1)-flavor
  QCD: Confronting Models with Lattice Studies}},
  \href{https://doi.org/10.1103/PhysRevD.81.074013}{\emph{Phys. Rev. D}
  {\bfseries 81} (2010) 074013}
  [\href{https://arxiv.org/abs/0910.5628}{{\ttfamily 0910.5628}}].

\bibitem{Abhishek:2017pkp}
A.~Abhishek, H.~Mishra and S.~Ghosh, \emph{{Transport coefficients in the
  Polyakov quark meson coupling model: A relaxation time approximation}},
  \href{https://doi.org/10.1103/PhysRevD.97.014005}{\emph{Phys. Rev. D}
  {\bfseries 97} (2018) 014005}
  [\href{https://arxiv.org/abs/1709.08013}{{\ttfamily 1709.08013}}].

\bibitem{Khaidukov:2018lor}
Z.~V. Khaidukov, M.~S. Lukashov and Y.~A. Simonov, \emph{{Speed of sound in the
  QGP and an SU(3) Yang-Mills theory}},
  \href{https://doi.org/10.1103/PhysRevD.98.074031}{\emph{Phys. Rev. D}
  {\bfseries 98} (2018) 074031}
  [\href{https://arxiv.org/abs/1806.09407}{{\ttfamily 1806.09407}}].

\bibitem{Khaidukov:2019icg}
Z.~V. Khaidukov and Y.~A. Simonov, \emph{{Thermodynamics of a quark-gluon
  plasma at finite baryon density}},
  \href{https://doi.org/10.1103/PhysRevD.100.076009}{\emph{Phys. Rev. D}
  {\bfseries 100} (2019) 076009}
  [\href{https://arxiv.org/abs/1906.08677}{{\ttfamily 1906.08677}}].

\bibitem{Mykhaylova:2020pfk}
V.~Mykhaylova and C.~Sasaki, \emph{{Impact of quark quasiparticles on transport
  coefficients in hot QCD}},
  \href{https://doi.org/10.1103/PhysRevD.103.014007}{\emph{Phys. Rev. D}
  {\bfseries 103} (2021) 014007}
  [\href{https://arxiv.org/abs/2007.06846}{{\ttfamily 2007.06846}}].

\end{thebibliography}\endgroup

\end{document}